\input phyzzx.tex
\tolerance=1000
\voffset=-0.0cm
\hoffset=0.7cm
\sequentialequations
\def\rl{\rightline}

\def\t1{{\tilde 1}}

\def\t{\theta}

\REF{\BEK}{J. Bekenstein, Phys. Rev. {\bf D23} (1981) 287; Phys. Rev. {\bf D49} (1994) 1912; Int. Journ. Thoer. Phys. {\bf 28} (1989) 967.}
\REF{\LAST}{E. Halyo, arXiv:0906.2164[hep-th].}
\REF{\BEKE}{J. Bekenstein, Lett. Nuov. Cimento {\bf 4} (1972) 737; Phys Rev. {\bf D7} (1973) 2333; Phys. Rev. {\bf D9} (1974) 3292.}
\REF{\HAW}{S. Hawking, Nature {\bf 248} (1974) 30; Comm. Math. Phys. {\bf 43} (1975) 199.}
\REF{\HOL}{G. 't Hooft, [arXiv:gr-qc/9310026].} 
\REF{\SUS}{L. Susskind, J. Math. Phys. {\bf 36} (1995) 6377, [arXiv:hep-th/9409089].}
\REF{\RAF}{R. Bousso, Rev. Mod. Phys. {\bf 74} (2002) 825, [arXiv:hep-th/0203101].}
\REF{\MAL}{J. Maldacena, Adv. Theo. Math. Phys. 2 (1998) 231, [arXiv:hep-th/9711200].}
\REF{\WIT}{E. Witten, Adv. Theo. Math. Phys. 2 (1998) 253, [arXiv:hep-th/9802150].}
\REF{\GKP}{S. S. Gubser, I. I. Klebanov and A. M. Polyakov, Phys. Lett. {\bf B428} (1998) 105, [arXiv:hep-th/9802109].}
\REF{\LEN}{L. Susskind, [arXiv:hep-th/9309145].}
\REF{\SBH}{E. Halyo, A. Rajaraman and L. Susskind, Phys. Lett. {\bf B392} (1997) 319, [arXiv:hep-th/9605112].}
\REF{\HRS}{E. Halyo, B. Kol, A. Rajaraman and L. Susskind, Phys. Lett. {\bf B401} (1997) 15, [arXiv:hep-th/9609075].}
\REF{\HP}{G. Horowitz and J. Polchinski, Phys. Rev. {\bf D55} (19997) 6189, [arXiv:hep-th/9612146];  Phys. Rev. {\bf D57} (1998) 2557, [arXiv:hep-th/9707170].}
\REF{\DAM}{T. Damour and G. Veneziano, Nucl. Phys. {\bf B568} (2000) 93, [arXiv:hep-th/9907030].}
\REF{\CHI}{D. Chialva, Nucl. Phys. {\bf B819} (2009) 256, arXiv:0903.3977[hep-th].}
\REF{\EDI}{E. Halyo, Int. Journ. Mod. Phys. {\bf A14} (1999) 3831, [arXiv:hep-th/9610068]; Mod. Phys. Lett. {\bf A13} (1998), [arXiv:hep-th/9611175].}
\REF{\DES}{E. Halyo, [arXiv:hep-th/0107169].}
\REF{\UNI}{E. Halyo, JHEP {\bf 0112} (2001) 005, [arXiv:hep-th/0108167].}
\REF{\ED}{E. Witten, Adv. Theo. Math. Phys. 2 (1998) 505, [arXiv:hep-th/9803131].}
\REF{\VER}{E. Verlinde, [arXiv:hep-th/0008140].}
\REF{\DBH}{E. Halyo, JHEP {\bf 0206} (2002) 012, [arXiv:hep-th/0201174].}
\REF{\MED}{A. Medved, [arXiv:hep-th/0201215].}

\singlespace
\rl{SU-ITP-09/31}
\pagenumber=0
\normalspace
\medskip
\bigskip
\titlestyle{\bf{Bekenstein Entropy is String Entropy}}
\smallskip
\author{ Edi Halyo{\footnote*{e--mail address: halyo@.stanford.edu}}}
\smallskip
\centerline {Department of Physics} 
\centerline{Stanford University} 
\centerline {Stanford, CA 94305}
\smallskip
\vskip 2 cm
\titlestyle{\bf ABSTRACT}

We argue that Bekenstein entropy can be interpreted as the entropy of an effective string with a rescaled tension. Using the AdS/CFT correspondence we show that the Bekenstein
entropy on the boundary CFT is given by the entropy of a string at the stretched horizon of the AdS black hole in the bulk. The gravitationally redshifted tension and energy of the
string match those required to reproduce Bekenstein entropy.

\singlespace
\vskip 0.5cm
\endpage
\normalspace

\centerline{\bf 1. Introduction}
\medskip

The Bekenstein bound[\BEK] on the entropy of a weakly gravitating system is given by 
$$S \leq S_{Bek}=2 \pi E R \eqno(1)$$ 
where $E$ and $R$ are the energy and size of the system respectively. The form of Bekenstein
entropy, $S_{Bek}$, is very suggestive; it looks like the entropy of an effective string with energy $E$ and string tension $T=1/2 \pi R^2$. This is very surprising because the bound applies
to nongravitational (or weakly gravitating) systems in which string theory should play no role. Moreover, the string in question needs to have an extremely small tension compared
to $1/ 2 \pi \ell_s^2$, the tension of a fundamental string since in general $R>>\ell_s$. It would be interesting to find out if such an effective string really exists. 
If it does, one would have to explain its relation to fundamental strings and the origin of its very small tension.

Recently, in ref. [\LAST], it was shown that there is a holographic relation between the Bekenstein entropy bound and the holographic entropy bound $S \leq A/4 G$[\BEKE,\HAW]. 
The former bound holds for nongravitational systems whereas the latter holds for strongly gravitating systems. Therefore, the two bounds may be related by holography[\HOL,\RAF]; 
the latter applies to the gravitational theory in the bulk whereas the former applies to
the field theory on the boundary. This idea can be realized by using the AdS/CFT correspondence[\MAL-\GKP] which is explicitly holographic. The Bekenstein entropy 
bound can be seen as a lower bound on the energy of a system with a given entropy and size. On the other hand, the entropy of an AdS black hole does not depend on the AdS radius.
Using these two facts it can be shown that the Bekenstein entropy bound is saturated on the boundary when the
boundary energy corresponding to an AdS black hole is minimized with respect to the AdS radius. This happens when the black hole and AdS radii are equal.

On the other hand, it has been known for some time that black holes and other black objects can be described by strings on their stretched horizons[\LEN-\UNI]. Such a string has an energy
that is much higher than the mass of the black hole whereas its entropy matches that of the black hole. 
To an asymptotic observer it looks like a string with a gravitationally redshifted tension which is much lower than $1/ 2 \pi \ell_s^2$. The asymptotic mass of the
string matches that of the black hole due to the redshift between the stretched horizon and infinity. Its entropy continues to match that of the black hole since entropy
does not redshift. In fact, it has been shown that such strings can also describe black holes in AdS and dS spaces.

In this paper, we would like to combine the results mentioned in the last two paragraphs and argue that the Bekenstein entropy can be seen as the entropy of a string that lives on the
stretched horizon of a black hole that holographically corresponds to the boundary field theory. Since the only explicit example of holography is the AdS/CFT correspondence,
we consider an AdS black hole and the corresponding thermal state of the boundary CFT. We show that the black hole can be described by a string on its stretched horizon and an observer
on the boundary sees this string with the correct redshifted tension and energy to describe the Bekenstein entropy.

\bigskip
\centerline{\bf 2. AdS Black Holes and Strings on the Stretched Horizon}
\medskip

We can investigate the possible relation between the Bekenstein entropy and an effective string by using the AdS/CFT correspondence. We first consider an AdS black hole and show that
its description in terms of the boundary CFT is intriguingly similar to that of an effective string with a rescaled tension. We then consider a string on the stretched horizon of the 
black hole and show that for an observer on the boundary it has exactly the entropy and tension of the effective string. The energy of the effective string is in general greater than the 
CFT energy. The origin of this mismatch is not clear to us. However, when
the Bekenstein bound is saturated, i.e. when the AdS and black hole radii are equal, the energy of the string matches that of the CFT making the relation Bekenstein and effective string 
entropies complete.

The AdS black hole solution (in any dimension $D$) is given by the metric
$$ds^2=-\left(1-{\mu^d \over r^d}+{r^2 \over L^2}\right) dt^2+ \left(1-{\mu^d \over r^d}+{r^2 \over L^2}\right)^{-1} dr^2 +r^2 d \Omega^2_{D-2} \eqno(2)$$
Here $\mu^d=16 \pi G_D M/(D-2)$ where $M$ is the mass of the black hole, $\Lambda=-(D-1)(D-2)/2L^2$ is the cosmological constant and $d=D-3$. The black hole horizon is at $r=R$ which
is the largest radius that satisfies
$$1-{\mu^d \over R^d}+{R^2 \over L^2}=0 \eqno(3)$$
The mass of the black hole can be written as
$$M=E_{AdS}={(d+1) \over {16 \pi G_D}} R^d \left({1+{R^2 \over L^2}}\right) \eqno(4)$$
with the Hawking temperature of the black hole
$$T_{AdS} = {1 \over {4 \pi R}} \left(d+ (d+2) {R^2 \over L^2}\right) \eqno(5)$$
The entropy of the black hole is given by the Bekenstein--Hawking formula
$$S_{AdS}= {R^{d+1} \over 4 G_D} \eqno(6)$$

On the boundary $S^{(D-2)}$ this black hole is described by a thermal state of a CFT[\WIT]. Following ref. [\VER] we use the conformal invariance of the boundary CFT to rescale the radius of
the $S^{(D-2)}$ to be $R$ rather than $L$. Thus we need to rescale all boundary energies by a factor of $L/R$. The boundary energy becomes
$$E_{CFT}={(d+1) \over {16 \pi G_D}}\left({R^{d+1} \over L} + {L R^{d-1}}\right) \eqno(7)$$
with the boundary temperature
$$T_{CFT}={1 \over {4 \pi L}} \left((d+2)+ d{L^2 \over R^2}\right) \eqno(8)$$
The entropy of the CFT is the same as the bulk black hole
$$S_{CFT}=S_{AdS}={R^{d+1} \over 4 G_D} \eqno(9)$$

In ref. [\VER] it was shown that we can view the energy in eq. (7) as a sum $E_{CFT}=E_E+E_C$ where
$$E_E={(d+1) \over {16 \pi G_D}} {R^{d+1} \over L} \eqno(10)$$
and 
$$E_C={(d+1) \over {16 \pi G_D}} {L R^{d-1}} \eqno(11)$$
$E_E$ and $E_C$ are the extensive gas and the subextensive Casimir contributions to the boundary energy respectively. The entropy can be written in a Cardy--like formula
for CFTs
$$S_{CFT}={4 \pi \over (d+1)} R \sqrt{E_C(E_{CFT}-E_C)} \eqno(12)$$
Using eqs. (10) and (11) we can write this entropy as
$$S_{CFT}={4 \pi L \over (d+1)} E_E \eqno(13)$$
This can be viewed as the entropy of an effective string with energy $E_{str}=2 \epsilon E_E/(d+1)$ and tension $T_{str}=\epsilon^2/2 \pi L^2$.
where we allowed for a parameter $\epsilon$ that may depend on $R,L,d$. Note that, in general, the energy of the effective string is not equal to $E_{CFT}$; in fact, for 
$R>L$ we find $2E_E>E_{CFT}$.
It seems that the effective string carries more energy than the CFT which is quite strange.
This may be a potential problem in general for $R>L$. However, we will see below that this problem is resolved when $R=L$ and the Bekenstein entropy bound is saturated.

Recently it was shown that the Bekenstein entropy bound on the boundary is holographically related to the Bekenstein--Hawking entropy bound in the bulk by using the AdS/CFT
correspondence[\LAST]. On the boundary of AdS space the Bekenstein entropy bound is modified to
$$S_{Bek}={ 2\pi \over (d+1)} E_{CFT}R \eqno(14)$$
For large AdS black holes with $R>L$ we find that $S_{CFT}<S_{Bek}$ and the bound is not saturated. This can be easily seen from eqs. (7) and (10) because $E_{CFT}R<2 E_E L$ for $R>L$.
Notice that the black hole entropy eq. (9) (and therefore the boundary entropy) is independent of the AdS radius $L$ whereas the boundary energy is not. 
On the other hand, the Bekenstein entropy 
bound can also be seen as a lower bound on the boundary energy for a given size and entropy. Therefore, in order to saturate the bound,
for a given $R$ and therefore fixed $S_{AdS}=S_{CFT}$ we need to minimize
$E_{CFT}$ with respect to the AdS radius $L$. The minimum is obtained for $L=R$ when the AdS radius equals the black hole radius. Then the Bekenstein entropy bound is saturated on the
boundary since for $L=R$ we find $E_{CFT}R=2 E_E L$ and $S_{Bek}=S_{CFT}$. 

The effective string that carries this entropy does not look like an fundamental string but one with energy $E_{str}= \epsilon E_{CFT}/(d+1)$ and tension 
$T_{str}= \epsilon^2/2 \pi L^2$. The question is whether there is such a string 
in the theory. The candidate for the effective string is a string that lives on the stretched horizon of the AdS black hole. It is known that many black holes and other black objects[\UNI]
(including de Sitter space[\DES]) can be described in terms of a long string located at the stretched horizon. We review this description for the AdS black hole below.

In order to find out the properties of the string at the stretched horizon we need to take the near horizon limit of the metric[\DBH,\MED].
Near the horizon we have $r=R+y$ with $y<<R$ and the metric becomes 
$$ds^2=-\left({y \beta \over L}\right) dt^2+ \left({y \beta \over L}\right)^{-1} dr^2+R^2 d \Omega^2_{D-2} \eqno(15)$$
where 
$$\beta=(d+2) {R \over L}+ d {L \over R} \eqno(16)$$
The proper distance to the horizon is given by
$$\rho= 2 \sqrt{{{yL} \over \beta}} \eqno(17)$$
In terms of the proper distance, the metric becomes
$$ds^2=-\left({\beta^2 \over {4L^2}}\right) \rho^2 dt^2+d\rho^2+ R^2 d \Omega_{D-2}^2 \eqno(18)$$
which is Rindler space as expected. The dimensionless Rindler time is 
$$\tau_R={\beta \over {2L}} t \eqno(19)$$
The dimensionless Rindler energy is obtained by using
$$[E_R,\tau_R]=[E_R,t]{\beta \over 2L}={dE_R \over dE}{\beta \over 2L}=1 \eqno(20)$$
which gives
$$2 \pi E_R = {R^{d+1} \over {4G_D}} \eqno(21)$$
which is the black hole entropy. This is expected since the Rindler temperature is $T_R=1/2 \pi$ and the Rindler entropy is the black hole entropy, $S_R=S_{CFT}$. Using the first law of
thermodynamics $dE=TdS$ we find
$$E_R={1 \over {2 \pi}} S_R={1\over {2 \pi}} S_{CFT} \eqno(22)$$
where we used eq. (21).

On the stretched horizon, the energy of the string (with tension $T=1/2 \pi \ell_s^2$) is $E_{sh}=E_R/\ell_s$ whereas its entropy is the same as the Rindler energy $S_{sh}=E_R=S_{AdS}$. 
The temperature of the
string is the Hagedorn temperature $T_{sh}=1/2 \pi \ell_s$. Note that this string is much more massive than the AdS black hole. Seen from the boundary of AdS
space this string will have a gravitationally redshifted energy and tension but the same entropy. The redshift factor between the stretched horizon and a cosmological observer (with 
time $t$) is $\ell_s \beta/2L$ obtained from eq. (19). In addition, there is the rescaling factor on the boundary which gives an additional factor of $L/R$. We find that the overall redshift
factor between the stretched horizon and the boundary is $\ell_s \beta/ 2R$. As a result, for an observer on the boundary the string temperature becomes
$$T={\beta \over {4 \pi R}}= {1 \over {4 \pi L}} \left((d+2)+ d{L^2 \over R^2}\right) \eqno(23)$$
which is precisely $T_{CFT}$. The entropy which is a number does not redshift so $S_{str}=S_{CFT}$ as expected.

Now let us find out the rescaled tension of the string as it is seen on the boundary. Taking into account the overall redshift factor we get
$$T_{str}={\beta^2 \over {8 \pi R^2}}= {1 \over {8 \pi L^2}} \left((d+2)+ d{L^2 \over R^2}\right)^2 \eqno(24)$$
From the above equation, we identify the factor $\epsilon=L \beta /2R$. When $R=L$, $\epsilon=d+1$. The energy of the string seen from the boundary is
$$E_{str}={{L \beta} \over {(d+1)R}}E_E={{2 \epsilon} \over (d+1)}E_E \eqno(25)$$
where we used eqs. (10) and (21). When the Bekenstein bound is saturated, $R=L$ and we find that the tension of the string becomes $T=(d+1)^2/ 2 \pi L^2$ matching the tension 
of the effective string.
The string energy becomes $E_{str}=2E_E$ which agrees with the energy of the effective string. Note that now $2E_E=E_{CFT}$ and the mismatch between the string and CFT energies 
disappears when the Bekenstein bound is saturated. Thus we find that the effective string we deduced from the boundary CFT expressions is a fundamental string on the stretched
horizon of the AdS black hole seen from the boundary.


\bigskip
\centerline{\bf 3. Conclusions and Discussion}
\medskip

In this paper we showed that the similarity between the Bekenstein entropy and the entropy of an effective string is not a coincidence. By using the AdS/CFT
correspondence, we showed that the effective string describes a string at the stretched horizon of the AdS black hole seen from the boundary. The AdS black hole is described by string
at its stretched horizon with a mass much larger than that of the black hole and an entropy that matches the black hole entropy. The Bekenstein entropy on the boundary is obtained
when the Bekenstein bound is saturated, i.e. when the black hole radius matches the AdS radius. We showed that for $R=L$, taking into account the gravitational
redshift between the stretched horizon and the boundary this string has the correct energy and tension to be the effective string that reproduces the Bekenstein entropy.

In fact, the string on the stretched horizon describes the AdS black hole for all $R>L$ and therefore the corresponding thermal states of the CFT which do not saturate the Bekenstein
entropy bound. As we saw above, for $R>L$, the string reproduces the temperature and entropy of the thermal CFT but not its energy. The redshifted string energy is greater than that 
of the CFT for $R>L$. Therefore, any thermal boundary CFT state can be described by an effective string but in these cases the string carries more energy than the CFT.
We do not understand the origin of this discrepancy; however we note that it disappears for $R=L$ exactly when the Bekenstein bound is saturated. 

Strictly speaking, we have only shown the connection between the Bekenstein entropy bound and string entropy for the case AdS black holes which correspond to thermal states of the
boundary CFT. Of course this is due to the
fact that the AdS/CFT correspondence is the only holographic setup in which this connection can be explicitly realized. Even in the context of the AdS/CFT correspondence it is hard to
find bulk states other than black holes with entropy that saturate the Bekenstein bound on the boundary.

\bigskip
\centerline{\bf Acknowledgements}

I would like to thank the Stanford Institute for Theoretical Physics for hospitality.


\vfill

\refout

\end
\bye